\documentclass[12pt,preprint]{aastex}
\shorttitle{Sun's Path Through the Galaxy}
\shortauthors{Gies \& Helsel}

\begin{document}

\received{2005 January 5}
\accepted{2005 March 14}

\title{Ice Age Epochs and the Sun's Path Through the Galaxy}

\author{D. R. Gies and J. W. Helsel}
\affil{Center for High Angular Resolution Astronomy and \\
 Department of Physics and Astronomy,\\
 Georgia State University, P. O. Box 4106, Atlanta, GA 30302-4106; \\
 gies@chara.gsu.edu, helsel@chara.gsu.edu}

\slugcomment{ApJ, in press}
\paperid{61945}

%%%%%%%%%%%%%%%%%%%%%%%%%%%%%%%%%%%%%%%%%%%%%%%%%%%%%%%%%%%%%%%

\begin{abstract}
We present a calculation of the Sun's motion through the 
Milky Way Galaxy over the last 500 million years.  The integration 
is based upon estimates of the Sun's current position and speed 
from measurements with {\it Hipparcos} and upon a realistic 
model for the Galactic gravitational potential.   We estimate 
the times of the Sun's past spiral arm crossings for 
a range in assumed values of the spiral pattern angular speed. 
We find that for a difference between the mean solar and pattern speed of 
$\Omega_\odot - \Omega_p = 11.9 \pm 0.7$ km~s$^{-1}$~kpc$^{-1}$
the Sun has traversed four spiral arms at times that appear to 
correspond well with long duration cold periods on Earth. 
This supports the idea that extended exposure to the higher 
cosmic ray flux associated with spiral arms can lead to increased 
cloud cover and long ice age epochs on Earth. 
\end{abstract}


%%%%%%%%%%%%%%%%%%%%%%%%%%%%%%%%%%%%%%%%%%%%%%%%%%%%%%%%%%%%%%%

\section{Introduction}                              % Section 1

Since its birth the Sun has made about 20 cycles around 
the Galaxy, and during this time the Sun has made many 
passages through the spiral arms of the disk.  There is a 
growing interest in determining how these passages may 
have affected Earth's environment.  Shaviv (2002, 2003) 
makes a persuasive argument that there is a correlation between 
extended cold periods on Earth and Earth's exposure to a
varying cosmic ray flux (CRF).  Shaviv proposes that the 
CRF varies as the Sun moves through Galactic spiral arms,
regions with enhanced star formation and supernova rates
that create more intense exposure to cosmic rays. 
The CRF experienced by Earth may affect the atmospheric 
ionization rate and, in turn, the formation of charged aerosols 
that promote cloud condensation nuclei \citep{har01,eic02}.
\citet{mar00} show that there is a close correlation between 
the CRF and low altitude cloud cover over a 15 year time span. 
Thus, we might expect that extended periods of high CRF lead to increased 
cloud cover and surface cooling that result in long term (Myr) ice ages. 
Spiral arm transits may affect Earth in other ways as well.
\citet{yeg04} suggest that during some spiral passages the 
Earth may encounter interstellar clouds of sufficient density 
to alter the chemistry of the upper atmosphere and 
trigger an ice age of relatively long duration.
The higher stellar density in the arms may more effectively 
perturb the Oort cloud of comets and lead to a greater 
chance of large impacts on Earth, and this combined with 
the possible lethal effects of nearby supernova explosions 
could cause mass extinctions during during passages through 
the spiral arms \citep{lei98}.  On the other hand, the 
record of terrestrial impact craters suggests a variation 
on a time scale shorter than the interarm crossing time, 
but possibly related to the Sun's oscillations above and 
below the disk plane \citep{sto98}.   

A comparison of the geological record of temperature variations
with estimates of the Sun's position relative to the spiral arms 
of the Galaxy is difficult for a number of reasons.  
First, our location within the disk makes it hard to discern the 
spiral structure of the Galaxy, particularly in more distant 
regions.  Nevertheless, there is now good evidence that 
a four-arm spiral pattern is successful in explaining the 
emissions from the star-forming complexes of the Galaxy
\citep{rus03}.  Second, the angular rotation speed of the 
Galactic spiral pattern is still poorly known, with estimates
ranging from 11.5 \citep{gor78} to 30 km~s$^{-1}$~kpc$^{-1}$ 
\citep*{fer01} (see reviews in \citealt{sha03}, \citealp*{bis03}, 
and \citealt{mar04}).  Finally, the Sun's orbit in the 
Galaxy is not circular, and we need to account for the Sun's 
variation in distance from Galactic center and in orbital speed 
to make an accurate estimate of the Sun's position in the past. 

Here we present such a calculation of the Sun's path through 
the Galaxy over the last 500~Myr.  It is based upon the Sun's
current motion relative to the local standard of rest as 
determined from parallaxes and proper motions from the {\it Hipparcos 
Satellite} \citep{dea98} and on a realistic model of the 
Galactic gravitational potential \citep{deb98}.  We discuss 
how the spiral pattern speed is critical to the estimates 
of the times of passage through the spiral arms, and we show
a plausible example that is consistent with the occurrence
of ice ages during spiral arm crossings. 

%%%%%%%%%%%%%%%%%%%%%%%%%%%%%%%%%%%%%%%%%%%%%%%%%%%%%%%%%%%%%%%

\section{Integration of the Sun's Motion}           % Section 2

An integration of the Sun's motion was made using a cylindrical coordinate
system for the Galaxy of $(R, \phi, Z)$.  We first determined the position and
resolved velocity components of the Sun in this system using the
velocity of the Sun with respect to the local standard of rest 
\citep{dea98} and the Sun's position relative to the plane \citep*{hol97}.  
We then performed integrations backward in time using a fourth-order 
Runge-Kutta method and a model for the Galactic potential from \citet{deb98}.
We adopted the model (\#2) from \citet{deb98} that uses a Galactocentric
distance of $R_o=8.0$~kpc and a disk stellar density exponential 
scale length of $R_{d\star}=2.4$~kpc.
This model has a circular velocity at $R_o =8.0$~kpc of 217.4 km~s$^{-1}$.
We used time steps of 0.01 Myr over a time span of 500 Myr. 
Note that the model potential is axisymmetric and does not account 
for the minor variations in the field near spiral arms. 
We also ignore accelerations due to encounters with giant molecular 
clouds, since their effect is small over periods less than 1~Gyr
(at least in a statistical sense; \citealt{jen92}). 
The full set of coordinates as a function of time is not included 
here, but interested readers can obtain the digital data from our web 
site\footnote{http://www.chara.gsu.edu/$^\sim$gies/solarmotion.dat}.

The Sun's journey in cylindrical coordinates is illustrated in 
Figure~1.   The top panel shows the temporal variation in distance
from Galactic center, and we see the radial oscillation that is
expected from the ``epicycle approximation'' for nearly circular 
orbits \citep{bin87}.  The period is 170 Myr and the corresponding 
frequency is 36.9 km~s$^{-1}$~kpc$^{-1}$, which is close to the expected 
value of $36.7\pm2.4$ km~s$^{-1}$~kpc$^{-1}$ based upon the local Oort 
constants \citep{fea97}.  The middle panel shows the advance in 
azimuthal position with the orbit (small departures from linearity 
reflect speed variations that conserve angular momentum).  
The Sun has completed just over two circuits of the Galaxy over this time span.  
The lower panel shows the oscillations above and below the Galactic plane.  
The period is approximately 63.6~Myr, but there are cycle to cycle variations 
caused by the varying radial density in the model.  This period $P$ is
approximately related to the mid-plane density at the average radius, 
$\rho = (26.43~{\rm Myr}/P)^2$ $M_\odot$~pc$^{-3}$ \citep{bin87}. 
The period for our model of the solar motion corresponds to a mid-plane density 
of 0.17 $M_\odot$~pc$^{-3}$, which is close to current estimates of
the Oort limit of $0.15\pm 0.01$ $M_\odot$~pc$^{-3}$ \citep{sto98}. 
Thus, while the estimates of motion of the Sun in the $Z$ direction are 
secure for the recent past, probable errors in the period of approximately
$7\%$ may accumulate to as much as half a cycle error in the timing 
of the oscillations 500~Myr ago.   The errors in the estimates of the 
Sun's current Galactic motions \citep{dea98} have only a minor impact on 
these trajectories.  For example, the error in the $V$ component of motion
amounts to a difference of only $3^\circ$ in $\phi$ over this 500~Myr 
time span.  

\begin{figure}
\begin{center}
{\includegraphics[angle=90,height=12cm]{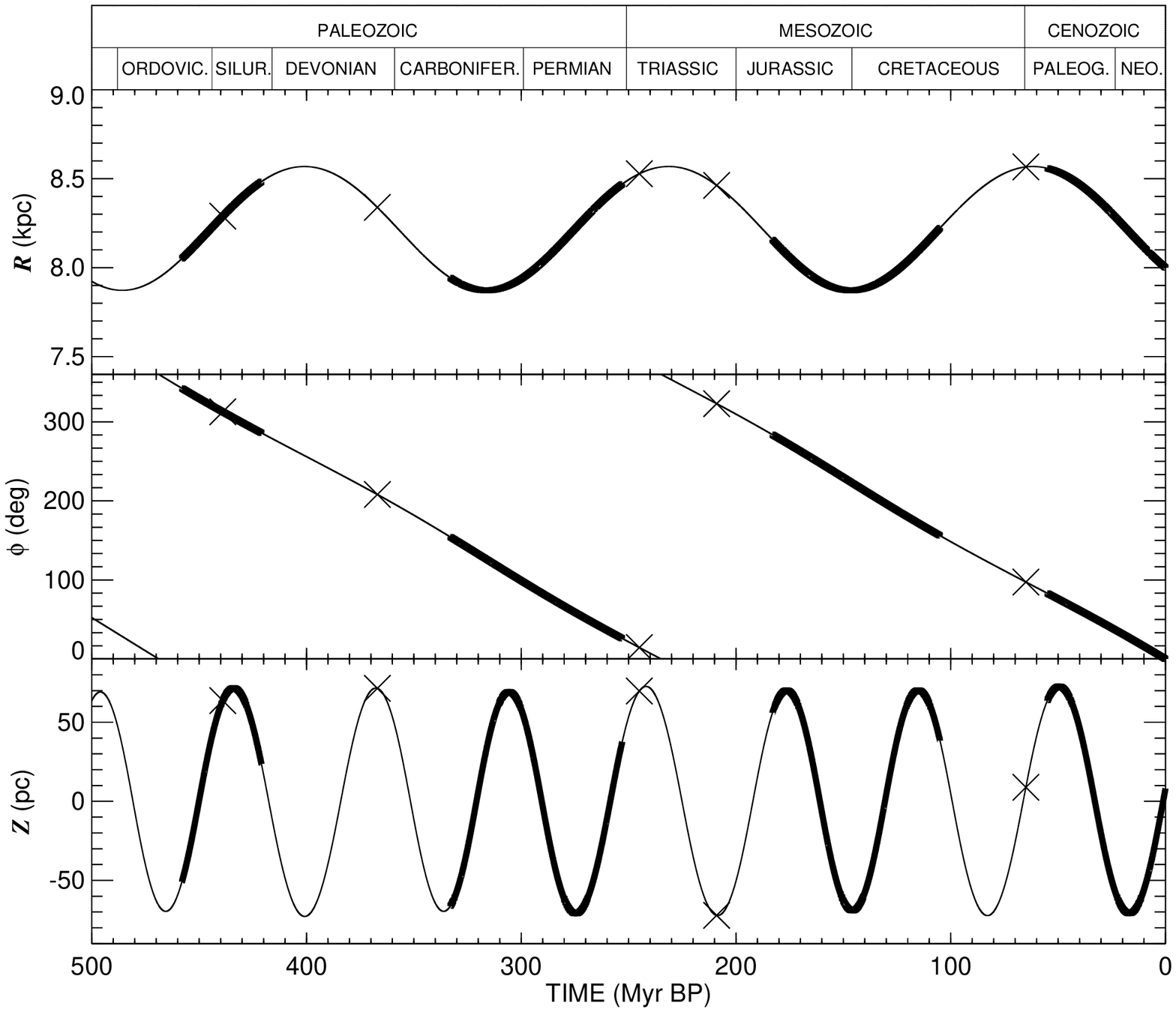}}
\end{center}
\caption{The Sun's position in the Galaxy over the last 500 Myr 
expressed in cylindrical coordinates, $R$ the distance from Galactic center
({\it top}), $\phi$ the azimuthal position in the disk relative to 
$\phi=0^\circ$ at present ({\it middle}), and $Z$ the distance from the 
plane ({\it bottom}).  Thick line portions mark icehouse epochs on 
Earth \citep{fra92}, and X signs indicate times of large mass 
extinctions on Earth.  The names of the geological eras and periods 
over this time span are noted at top.}
\label{fig1}
\end{figure}

\placefigure{fig1}     % Figure 1 - (R, phi, Z) motions

We next consider the motion of the Sun in the plane of the Galaxy 
relative to the spiral arm pattern.  The disk of the Galaxy
from the solar circle out-wards appears to display a four-arm 
spiral structure as seen in the emission of atomic hydrogen \citep*{bli83} 
and molecular CO \citep*{dam01} and in the distribution of 
star forming regions \citep{rus03}.  We show in Figure~2 the 
appearance of the Galactic spiral arm patterns based on the model of
\citet{wai92} but with some revisions introduced by 
\citet{cor03}\footnote{http://astrosun2.astro.cornell.edu/$^\sim$cordes/NE2001/}. 
This representation is very similar to the pattern adopted by \citet{rus03}.
We have rescaled the pattern from a solar Galactocentric radius of 8.5 kpc
to a value of 8.0 kpc for consistency with our model of Galactic 
potential from \citet{deb98}.  Each arm is plotted with an assumed width of 
0.75 kpc \citep{wai92} and each is named in accordance with the 
scheme of \citet{rus03}.   The dotted line through the center of the 
Galaxy indicates the current location of the central bar according to \citet{bis03}.
The pattern speed of the bar may be similar to that of the arms \citep{iba95} or  
it may be faster than that of the arms \citep{bis03}, in which case the bar -- arm 
relative orientation will be different in the past. 

\begin{figure}
\begin{center}
\plotone{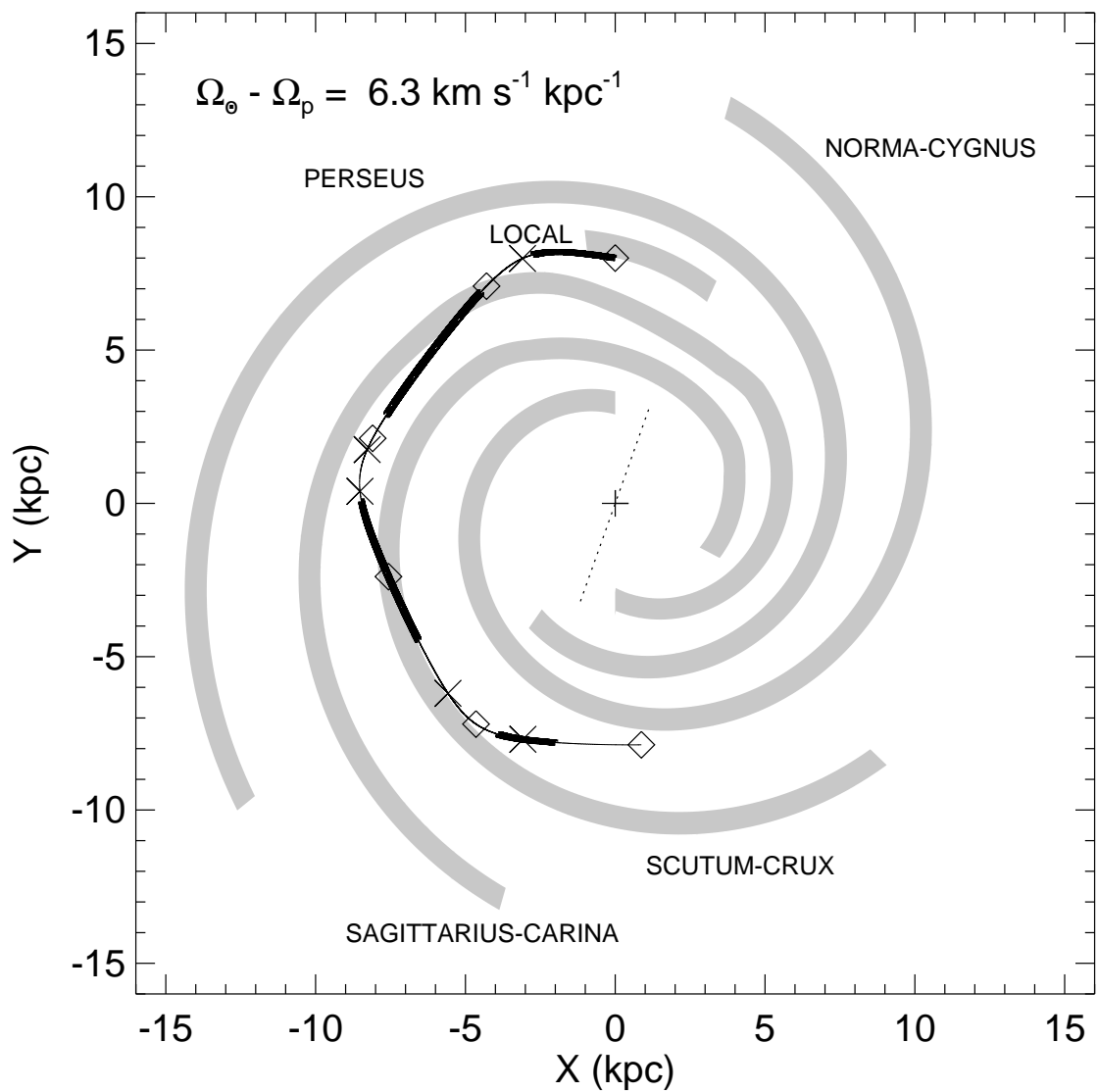}
\end{center}
\caption{A depiction of the spiral arm pattern of the Galaxy as viewed
from above the plane.  The plus sign marks the center of the Galaxy 
while the main four arms plus the local (Orion) spur are indicated 
as gray shaded regions.  The dotted line through the center of the Galaxy
indicates the location of the central bar \citep{bis03}.  The Sun's
path in the reference frame of the spiral arms is indicated with 
a solid line (for $\Omega_p=20$ km~s$^{-1}$~kpc$^{-1}$), 
and diamonds mark time intervals of 100 Myr back in time
from the present ({\it top diamond}).  The thick portions correspond
to icehouse times and the X signs indicate times of large mass extinctions.}
\label{fig2}
\end{figure}

\placefigure{fig2}     % Figure 2 - Motion wrt arms (Omega_p=20)

The placement of the Sun's trajectory in this diagram depends critically 
on the relative angular pattern speeds of the Sun and the spiral arms.
The mean advance in azimuth in our model of the Sun's motion corresponds 
to a solar angular motion of $\Omega_\odot = 26.3$ km~s$^{-1}$~kpc$^{-1}$.  
If the difference in the solar and spiral arm pattern speeds, 
$\Omega_\odot-\Omega_p$, is greater than zero, then the Sun overtakes 
the spiral pattern and progresses in a clockwise direction in our 
depiction of the Galactic plane.  Unfortunately, the spiral pattern 
speed is not well established and may in fact be different in the 
inner and outer parts of the Galaxy \citep{sha03}.   
Several recent studies \citep{ama97,bis03,mar04} advocate a 
spiral pattern speed of $\Omega_p=20\pm5$ km~s$^{-1}$~kpc$^{-1}$, 
and we show in Figure~2 the Sun's trajectory projected onto the plane for this value
($\Omega_\odot-\Omega_p = 6.3$ km~s$^{-1}$~kpc$^{-1}$). 
Diamonds along the Sun's track indicate its placement at intervals of 100~Myr. 
We see that for this assumed pattern speed the Sun has passed through only 
two arms over the last 500~Myr.   However, if we assume a lower but still 
acceptable pattern speed of $\Omega_p=14.4$ km~s$^{-1}$~kpc$^{-1}$ 
(shown in Fig.~3 for $\Omega_\odot-\Omega_p = 11.9$ km~s$^{-1}$~kpc$^{-1}$), 
then the Sun has crossed four spiral arms in the past
500~Myr and has nearly completed a full rotation ahead of the spiral pattern. 
Thus, the choice of the spiral pattern speed dramatically influences
any conclusions about the number and timing of Sun's passages through 
the spiral arms over this time interval.  

\begin{figure}
\begin{center}
\plotone{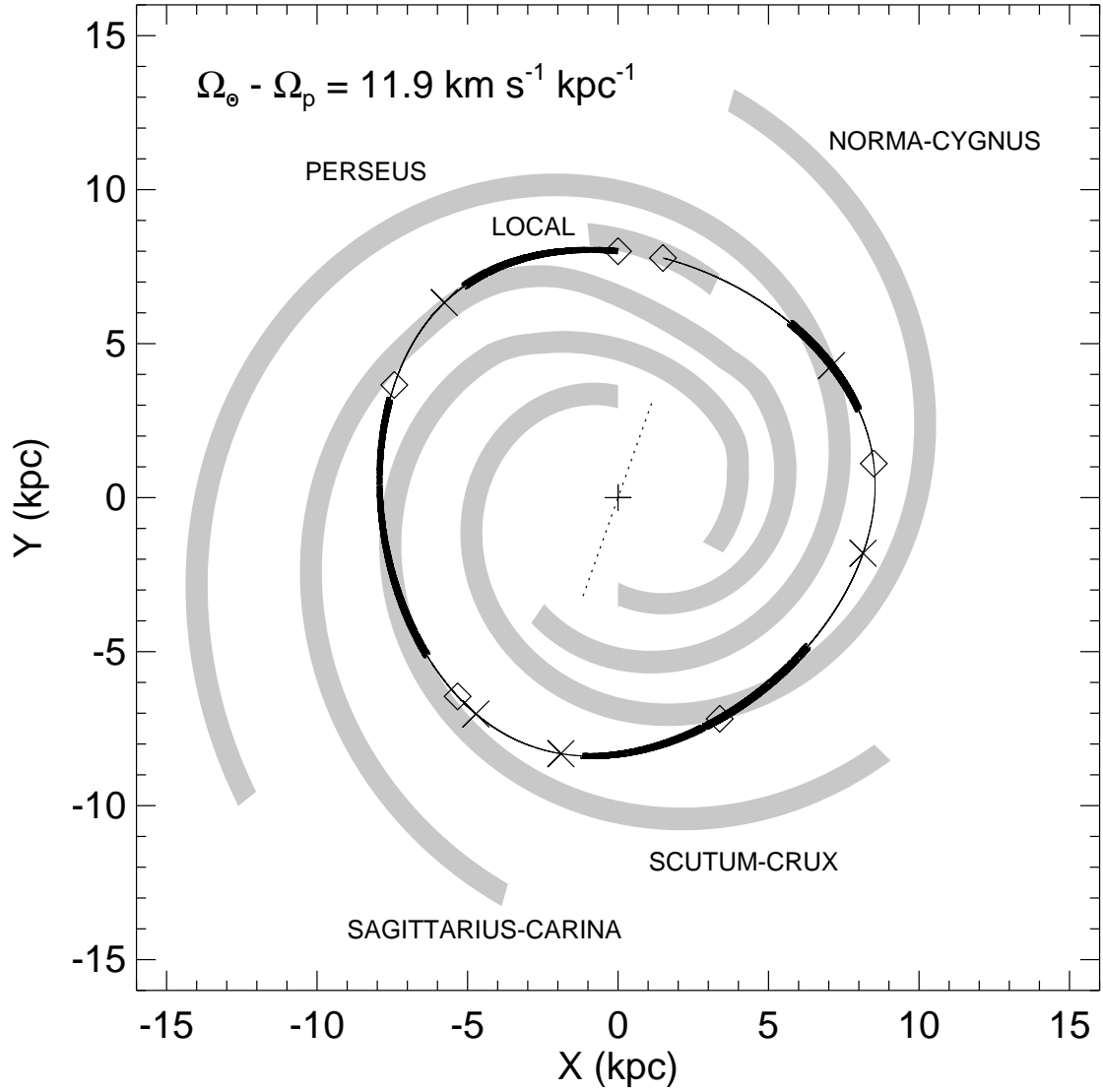}
\end{center}
\caption{A depiction of the Sun's motion relative to the spiral arm 
pattern in the same format as Fig.~2, but this time for a smaller 
spiral pattern speed ($\Omega_p=14.4$ km~s$^{-1}$~kpc$^{-1}$).}
\label{fig3}
\end{figure}

\placefigure{fig3}     % Figure 3 - Motion wrt arms (Omega_p=15)

The duration of a coherent spiral pattern is an open question, 
but there is some evidence that long-lived spiral patterns may 
be more prevalent in galaxies with a central bar.  
For example, numerical simulations of the evolution of barred spirals  
by \citet{rau99} suggest that spiral patterns may last 
several gigayears.  Their work suggests that the shortest time scale 
for the appearance or disappearance of a spiral arm is about 
1~Gyr.  Therefore, it is reasonable to assume that the 
present day spiral structure has probably been more or less intact 
over the last 500~Myr (at least in the region of the solar circle). 

%%%%%%%%%%%%%%%%%%%%%%%%%%%%%%%%%%%%%%%%%%%%%%%%%%%%%%%%%%%%%%%

\section{Discussion}                                % Section 3

\citet{sha03} argues that the Earth has experienced four large scale 
cycles in the CRF over the last 500~Myr (with similar cycle times 
back to 1 Gyr before the present).   Shaviv shows that the 
CRF exposure ages of iron meteorites indicate a periodicity of 
$143\pm10$~Myr in the CRF rate.  Since the cosmic ray production 
is related to supernovae and since Type~II supernovae will be more 
prevalent in the young star forming regions of the spiral arms, 
Shaviv suggests that the periodicity corresponds to 
the mean time between arm crossings (so that Earth has made four arm 
crossings over the last 500~Myr).  \citet{sha03} and \citet{sav03} show 
how the epochs of enhanced CRF are associated with cold periods on Earth.  
The geological record of climate-sensitive sedimentary layers 
(glacial deposits) and the paleolatitudinal distribution of 
ice rafted debris \citep*{fra92,cro99} indicate that the 
Earth has experienced periods of extended cold 
(``icehouses'') and hot temperatures (``greenhouses'')
lasting tens of million years \citep{fra92}.  
The long periods of cold may be punctuated by much more 
rapid episodes of ice age advances and declines \citep{imb92}.
The climate variations indicated by the geological evidence
of glaciation are confirmed by measurements of ancient 
tropical sea temperatures through oxygen isotope levels in 
biochemical sediments \citep{vei00}.  All of these studies 
lead to a generally coherent picture in which four periods of extended cold 
have occurred over the last 500~Myr, and the midpoints of these
ice age epochs (IAE) are summarized in Table~1 \citep[see][]{sha03}.
The icehouse times according to \citet{fra92} are indicated by 
the thick line segments in each of Figures 1, 2, and 3. 

\placetable{tab1}      % Table 1 - Midpoints of Ice Age Epochs
 
If these IAE do correspond to the Sun's passages through spiral 
arms, then it is worthwhile considering what spiral pattern speeds
lead to crossing times during ice ages.  We calculated the 
crossing times for a grid of assumed values of $\Omega_\odot - \Omega_p$
and found the value that minimized the $\chi^2_\nu$ residuals 
of the differences between the crossing times and IAE. 
There are two major error sources in the estimation of 
the timing differences.  First, the calculated arm crossing 
times depend sensitively on the placement of the spiral arms, 
and we made a comparison between the crossing times for our
adopted model and that of \citet{rus03} to estimate the 
timing error related to uncertainties in the position of the
spiral arms (approximately $\pm8$~Myr except in the case of the 
crossing of the Scutum--Crux arm on the far side of the Galaxy 
where the difference is $\approx 40$~Myr).  Secondly, there are errors 
associated with the estimated mid-times of the IAE, and we 
used the scatter between the various estimates in columns 2 -- 5
of Table~1 to set this error (approximately $\pm14$~Myr). 
We adopted the quadratic sum of these two errors in evaluating
the $\chi^2_\nu$ statistic of each fit.  The results of the 
fitting procedure for various model and sample assumptions 
are listed in Table~2.

\placetable{tab2}      % Table 2 - Arm crossing fits

The first trial fit was made by finding the $\chi^2_\nu$ minimum 
that best matched the crossing times with the IAE midpoints 
from \citet{sha03} (given in column 5 of Table~1 and noted as 
``Midpoint'' in column 2 of Table~2).  All four arm crossings 
were included in the calculation (indicated as 1 -- 4 in column 3
of Table~2) that used the adopted model for the Galactic 
potential with a Galactocentric distance $R_o = 8.0$~kpc and
and a stellar disk exponential scale length of $R_{d\star}=2.4$~kpc
(model \#2 from \citealt{deb98}; see columns 4 and 5 of Table~2). 
The best fit difference (column 6 of Table~2) is obtained 
with $\Omega_\odot - \Omega_p = 12.3 \pm 0.8$ km~s$^{-1}$~kpc$^{-1}$,
where the error was estimated by finding the limits for 
which $\chi^2_\nu$ increased by 1.  This fit gave reasonable 
agreement between the IAE and crossing times for all but the 
most recent crossing of the Sagittarius -- Carina arm.
Thus, we made a second fit (\#2 in Table~2) using only 
the crossings associated with IAE 2 -- 4, and this solution 
(with $\Omega_\odot - \Omega_p = 11.9 \pm 0.7$ km~s$^{-1}$~kpc$^{-1}$)
is the one illustrated in Figure~3.  The crossing times (given 
in the final column of Table~1) agree well with the adopted IAE
midpoints.   Our results are similar to the estimate of 
$\Omega_\odot - \Omega_p = 10.4 \pm 1.5$ km~s$^{-1}$~kpc$^{-1}$
from \citet{sha03} who assumed a circular orbit for the Sun in 
the Galaxy. 

We also computed orbits using two other models for the Galactic 
potential from \citet{deb98} and determined the best fit 
spiral speeds for these as well.  Fit \#3 in Table~2 was 
made assuming a larger Galactocentric distance $R_o = 8.5$~kpc
but with the same ratio of $R_{d\star}/R_o$ (model \#2b in 
\citealt{deb98}), and the resulting best fit spiral speed is 
the same within errors as that for our adopted model. 
We also computed an orbit for a potential with a larger value
of disk exponential scale length $R_{d\star}/R_o$ 
(model \#3 in \citealt{deb98}), but again the best fit spiral 
speed (fit \#4 in Table~2) is the same within errors as that 
for our adopted model.  Thus, the details of the adopted 
Galactic potential model have little influence on the derived
spiral pattern speed needed to match the IAE times. 

We might expect that the IAE midpoint occurs somewhat 
after the central crossing of the arm.  For example, 
\citet{sha03} suggests that the IAE midpoint may occur some 
21 -- 35~Myr after the central arm crossing due to the 
difference in the stellar and pattern speeds (so that the 
cosmic rays move ahead of arms as the stellar population does)
and to the time delay between stellar birth and supernova 
explosion of the SN~II cosmic ray sources.  Furthermore, 
if ice ages are triggered by encounters with dense clouds 
as suggested by \citet{yeg04}, then the ice age may not begin 
until the Sun reaches the gas density maximum at the center 
of the arm.  Thus, we calculated a second set of best fit spiral 
speeds to match the mean crossing and icehouse starting times 
\citep{fra92}, and these are listed as fits \#5 and \#6 in Table~2. 
This assumption leads to somewhat smaller values of 
$\Omega_\odot - \Omega_p$, but ones that agree within errors 
with all the other estimates. 

We offer a few cautionary notes about possible systematic errors
in this analysis.  First, the fit of the IAE and arm crossing times 
depends on the difference $\Omega_\odot - \Omega_p$, and if our 
assumed value of $\Omega_\odot$ eventually needs revision, then so 
too will the spiral pattern speed $\Omega_p$ need adjustment.  
For example, \citet{rei04} derive an angular rotation speed of 
$\Omega_{LSR} = 29.5\pm 1.9$ km~s$^{-1}$~kpc$^{-1}$ 
for the local standard of rest based upon 
Very Long Baseline Array observations of the proper motion
of Sgr~A$^\star$ with respect to two extragalactic radio sources. 
If we suppose the local Galactic rotation curve is flat, 
then $\Omega_\odot = \Omega_{LSR} ~8.0 / R_g = 28.7\pm 1.8$
km~s$^{-1}$~kpc$^{-1}$, where $R_g=8.23$~kpc is the Sun's mean 
Galactocentric distance. 
Adopting this value results in a spiral pattern speed of 
$\Omega_p = 16.8\pm 2.0$.  Second, our calculation ignores 
any orbital perturbations caused by close encounters with 
giant molecular clouds that cause an increase in the Sun's motion 
with respect to a circularly rotating frame of reference.  
\citet{nor04} present of a study of the ages and velocities of 
Galactic disk stars that indicates a net increase in the random 
component of motion proportional to time raised to the exponent 0.34.  
Thus, we would expect that the Sun's random speed has increased 
through encounters by only $\approx 4\%$ over the last 500~Myr, 
too small to change the orbit or the arm crossing times estimates 
significantly.  Third, we have ignored the deviations in the 
gravitational potential caused by the arms themselves.  
The Sun presumably slows somewhat during the arm crossings so 
that the duration of the passage is longer than indicated in 
our model, but since our model of the gravitational potential
represents an azimuthal average, the derived orbital period and 
interarm crossing times should be reliable.   

\citet{lei98} argue that mass extinctions may also preferentially 
occur during spiral arm crossings.  However, they proposed that 
a spiral pattern speed of $\Omega_p=19$ km~s$^{-1}$~kpc$^{-1}$
is required to find consistency between times of mass extinctions 
and spiral arm crossings, and if correct, then the relationship 
between ice ages and arm crossings would apparently be ruled out 
because $\Omega_p=19$ km~s$^{-1}$~kpc$^{-1}$ is too large for 
the inter-arm crossing time to match the intervals between IAE
(see Fig.~2 and Fig.~3).   We show the times of the five major 
mass extinctions as X signs in Figures 1 -- 3 
\citep{rau86,ben95,mat96}.   We see that in fact the 
lower value of $\Omega_p=14.4$ km~s$^{-1}$~kpc$^{-1}$ 
($\Omega_\odot - \Omega_p = 11.9$ km~s$^{-1}$~kpc$^{-1}$,
as shown in Fig.~3) also leads to a distribution of mass extinction 
times that fall close to or within a spiral arm passage, so the 
association of mass extinctions with arm crossings may also 
be viable in models with pattern speeds that are consistent with 
the ice age predictions.  

Our calculation of the Sun's motion in the Galaxy appears to 
be consistent with the suggestion that ice age epochs occur 
around the times of spiral arm passages as long as the spiral 
pattern speed is close to $\Omega_p=14 - 17$ km~s$^{-1}$~kpc$^{-1}$. 
However, this value is somewhat slower than the $20\pm5$ km~s$^{-1}$~kpc$^{-1}$
preferred in recent dynamical models of the Galaxy 
\citep{ama97,bis03,mar04}.  The resolution of this dilemma 
may require more advanced dynamical models that can accommodate 
differences between pattern speeds in the inner and outer parts
of the Galaxy (for example, a possible resonance between 
the four-armed spiral pattern moving with $\Omega_p=15$ km~s$^{-1}$~kpc$^{-1}$
and a ``two-armed'' inner bar moving with $\Omega_p=60$ km~s$^{-1}$~kpc$^{-1}$; 
\citealt{bis03}). 

%%%%%%%%%%%%%%%%%%%%%%%%%%%%%%%%%%%%%%%%%%%%%%%%%%%%%%%%%%%%%%%

\acknowledgments

We thank Walter Dehnen for sending us his code 
describing the Galactic gravitational potential. 
We also thank the referee and our colleagues Beth Christensen, 
Crawford Elliott, and Paul Wiita for comments on this work. 
Financial support was provided by the National Science 
Foundation through grant AST$-$0205297 (DRG).
Institutional support has been provided from the GSU College
of Arts and Sciences and from the Research Program Enhancement
fund of the Board of Regents of the University System of Georgia,
administered through the GSU Office of the Vice President
for Research.  

%%%%%%%%%%%%%%%%%%%%%%%%%%%%%%%%%%%%%%%%%%%%%%%%%%%%%%%%%%%%%%%

% References

%%%%%%%%%%%%%%%%%%%%%%%%%%%%%%%%%%%%%%%%%%%%%%%%%%%%%%%%%%%%%%%

% Figure captions

\input{epsf}

% Figure 1

% Figure 2

% Figure 3

%%%%%%%%%%%%%%%%%%%%%%%%%%%%%%%%%%%%%%%%%%%%%%%%%%%%%%%%%%%%%%%

% Tables

\clearpage

% Table 1
\begin{deluxetable}{lccccc}
\tabletypesize{\scriptsize}
\tablewidth{0pt}
\tablenum{1}
\tablecaption{Mid-points of Ice Age Epochs\label{tab1}}
\tablehead{
\colhead{Ice Age}              &  
\colhead{Crowell (1999)}       & 
\colhead{Frakes et al. (1992)} & 
\colhead{Veizer et al. (2000)} & 
\colhead{Shaviv (2003)}        &
\colhead{Arm Crossing (Fit \#2)}   \\
\colhead{Epoch}                &  
\colhead{(Myr BP)}             & 
\colhead{(Myr BP)}             &
\colhead{(Myr BP)}             &
\colhead{(Myr BP)}             &
\colhead{(Myr BP)}             }
\startdata
1\dotfill &   $<22$ &   $<28$ &\phn30 &\phn20 &\phn80 \\
2\dotfill &\phn155  &\phn144  &   180 &   160 &   156 \\
3\dotfill &\phn319  &\phn293  &   310 &   310 &   310 \\
4\dotfill &\phn437  &\phn440  &   450 &   446 &   446 \\
\enddata
\end{deluxetable}

% Table 2
\begin{deluxetable}{lccccc}
\tablewidth{0pt}
\tablenum{2}
\tablecaption{Fits of Spiral Arm Pattern Speed\label{tab2}}
\tablehead{
\colhead{Fit}    &
\colhead{IAE}    &  
\colhead{IAE}    & 
\colhead{$R_o$}  & 
\colhead{$R_{d\star}$} & 
\colhead{$\Omega_\odot-\Omega_p$} \\ 
\colhead{Number} &  
\colhead{Times} & 
\colhead{Sample} &
\colhead{(kpc)}  &
\colhead{(kpc)}  &
\colhead{(km s$^{-1}$ kpc$^{-1}$)} }
\startdata
1\dotfill & Midpoint & 1 -- 4 & 8.0 & 2.40 & $12.3 \pm 0.8$ \\
2\dotfill & Midpoint & 2 -- 4 & 8.0 & 2.40 & $11.9 \pm 0.7$ \\
3\dotfill & Midpoint & 2 -- 4 & 8.5 & 2.55 & $11.8 \pm 0.6$ \\
4\dotfill & Midpoint & 2 -- 4 & 8.0 & 2.80 & $11.8 \pm 0.7$ \\
5\dotfill & Starting & 1 -- 4 & 8.0 & 2.40 & $11.6 \pm 0.8$ \\
6\dotfill & Starting & 2 -- 4 & 8.0 & 2.40 & $11.4 \pm 0.6$ \\
\enddata
\end{deluxetable}

% Old Table 2
%\begin{deluxetable}{lccc}
%\tablewidth{0pt}
%\tablenum{2}
%\tablecaption{Spiral Arm Crossings Times for Several Pattern Speeds\label{tab2}}
%\tablehead{
%\colhead{}                                     &
%\multispan{3}  
%{Crossing Time For $\Omega_p$ (in km s$^{-1}$ kpc$^{-1}$)} \\ 
%\noalign{\smallskip}
%\cline{2-4}
%\noalign{\smallskip}
%\colhead{}              &  
%\colhead{$\Omega_p=14$} & 
%\colhead{$\Omega_p=15$} & 
%\colhead{$\Omega_p=16$} \\ 
%\colhead{Spiral Arm Name}                      &  
%\colhead{(Myr BP)}                             & 
%\colhead{(Myr BP)}                             &
%\colhead{(Myr BP)}                             }
%\startdata
%Sagittarius -- Carina\dotfill &\phn78 &\phn83 &\phn88 \\
%Scutum -- Crux       \dotfill &   151 &   165 &   210 \\
%Norma -- Cygnus      \dotfill &   302 &   323 &   382 \\
%Perseus              \dotfill &   436 &   459 &   488 \\
%\enddata
%\end{deluxetable}

%%%%%%%%%%%%%%%%%%%%%%%%%%%%%%%%%%%%%%%%%%%%%%%%%%%%%%%%%%%%%%

\end{document}